\documentclass[num-refs,twocolumn]{wiley-article}

\usepackage{hyperref}
\hypersetup{
    colorlinks=true,
    linkcolor=blue,
    filecolor=magenta,      
    urlcolor=cyan,
    citecolor=magenta
}

\definecolor{winter}{rgb}{0.85,0.08,0.2}
\definecolor{summer}{rgb}{0.95,0.53,0.18}
\definecolor{spring}{rgb}{0,0.8,0.6}

\usepackage{siunitx}

\papertype{Review}

\title{
Protein sequence-to-structure
learning: Is this the end(-to-end revolution)?
\\
{\Large Protein sequence-to-structure learning}
}

\author[1\authfn{1}]{Elodie Laine}
\author[2\authfn{1}]{Stephan Eismann}
\author[3\authfn{1}]{Arne Elofsson}
\author[4\authfn{1}]{Sergei Grudinin}

\contrib[\authfn{1}]{Equally contributing authors.}
\usepackage[final]{changes}

\affil[1]{Sorbonne Universit\'e, CNRS, IBPS,  Laboratoire de Biologie Computationnelle et Quantitative (LCQB), 75005 Paris, France}
\affil[2]{Dep. of Computer Science and Applied Physics, Stanford University, Stanford, CA 94305, USA} %
\affil[3]{Dep. of Biochemistry and Biophysics and Science for Life Laboratory, Stockholm University, Box 1031, 171 72 Solna, Sweden} 
\affil[4]{Univ. Grenoble Alpes, CNRS, Grenoble INP, LJK, 38000 Grenoble, France}
\corraddress{Sergei Grudinin, Univ. Grenoble Alpes, CNRS, Grenoble INP, LJK, 38000 Grenoble, France}
\corremail{sergei.grudinin@univ-grenoble-alpes.fr}

\fundinginfo{
EL was funded by the French national research agency grant ANR-17-CE12-0009.
SE was supported by a Stanford Bio-X Bowes Fellowship. 
AE was funded by grants from the Swedish, E-science Research Center, Swedish National Infrastructure for Computing, and Swedish Natural Science Research Council No VR-NT 2016-03798. 
}

\runningauthor{Laine et al.}

\begin{document}

\begin{frontmatter}
\maketitle

\begin{abstract}

The potential of deep learning has been recognized in the protein structure prediction community for some time, and became indisputable after CASP13. 
In CASP14, deep learning has boosted the field to unanticipated levels reaching near-experimental accuracy.
This success comes from advances transferred from other machine learning areas, 
as well as methods specifically designed to deal with protein sequences and structures, and their abstractions. 
Novel emerging approaches include
(i) geometric learning, {\it i.e.} learning on representations such as graphs, 3D Voronoi tessellations, and point clouds;
(ii) pre-trained protein language models leveraging attention;
(iii) equivariant architectures preserving the symmetry of 3D space;
(iv) use of large meta-genome databases;
(v) combinations of protein representations;
(vi) and finally truly end-to-end architectures, {\it i.e.} differentiable models starting from a sequence and returning a 3D structure.
Here, we provide an overview and our opinion of the novel deep learning approaches developed in the last two years and widely used in CASP14.

\end{abstract}

\end{frontmatter}

\section{Introduction}

In December 2020, the fourteenth edition of CASP marked a big leap in protein three-dimensional (3D) structure prediction. Indeed, deep learning-powered approaches have reached unprecedented levels of near-experimental accuracy. 
This achievement has been made possible thanks to the latest improvements in geometric learning and natural language processing (NLP) techniques, and to the amounts of sequence and structure data accessible today. 
\added{The fundamental basis for the revolution in structure prediction comes from the use of co-evolution. While traditional measures of co-variations in natural sequences led to a few successes \cite{Benner1877385,Gobel8208723,Ortiz10526366}, major improvements came from recasting the problem as an inverse Potts model \cite{Lapedes1999,Giraud11969452}. These ideas started to show their full potential about 10 years ago with the development of efficient methods dealing with large scale multiple sequence alignments} \cite{thomas2005graphical,Weigt19116270,Balakrishnan21268112}. 
\replaced{They}{These methods} enabled the modelling of 3D structures for large protein
families \cite{Morcos22106262,Taylor22102360,Sadowski22000804,Marks22163331,Sulkowska2012,marks2012protein}. 


Shifting from unsupervised statistical inference to supervised deep learning further boosted the accuracy of the predicted contacts, and extended the applicability of this conceptual framework to families with fewer sequences \cite{Skwark25375897,Wang_2017_PLoS} and to the prediction of residue-residue distances \cite{Xu31399549,Senior31942072}. 
These advances have significantly increased the protein structure modelling coverage of genomes \cite{Ovchinnikov26335199,Lamb30796988,Greener31484923}, and also of bacterial interactomes \cite{Hopf2014,Green33654096,Cong31296772}. Over the past years, the CASP community has contributed to these efforts, with an increasing number of teams developing and applying deep learning approaches. 

The emergence of novel deep learning techniques has inspired a re-visit of the representations best suited for biological objects (protein sequences and structures).
In particular, advances in the treatment of language \cite{vaswani2017attention} and of 3D geometry \cite{Kondor2013,cohen2016steerable,gilmer17a,thomas2018tensor,bronstein2021geometric} by deep learning architectures have further benefited the field of protein structure and function prediction. 
Expanding on this progress, the DeepMind team demonstrated in CASP14 that it is possible to produce extremely accurate 3D models of proteins by learning end-to-end from sequence alignments of related proteins \cite{jumper2021highly}. This implies being able to capture long-range dependencies between amino acid residues, to transform these dependencies into structural constraints, and to preserve the symmetry and properties of the 3D space when operating on protein structures. 

This article is a follow-up to \citet{kandathil2019recent}. It aims at providing CASP participants and observers with some overview of the recent developments in deep learning applied to protein structure prediction, and some comprehensive description of key concepts we think have contributed to the formidable improvements we have witnessed in the latest CASP edition. We then discuss the implications of these improvements, the next-to-solve problems, and speculate about the future of structural (and computational) biology.

\begin{figure}[ht!]
\centering
\includegraphics[width=1\columnwidth]{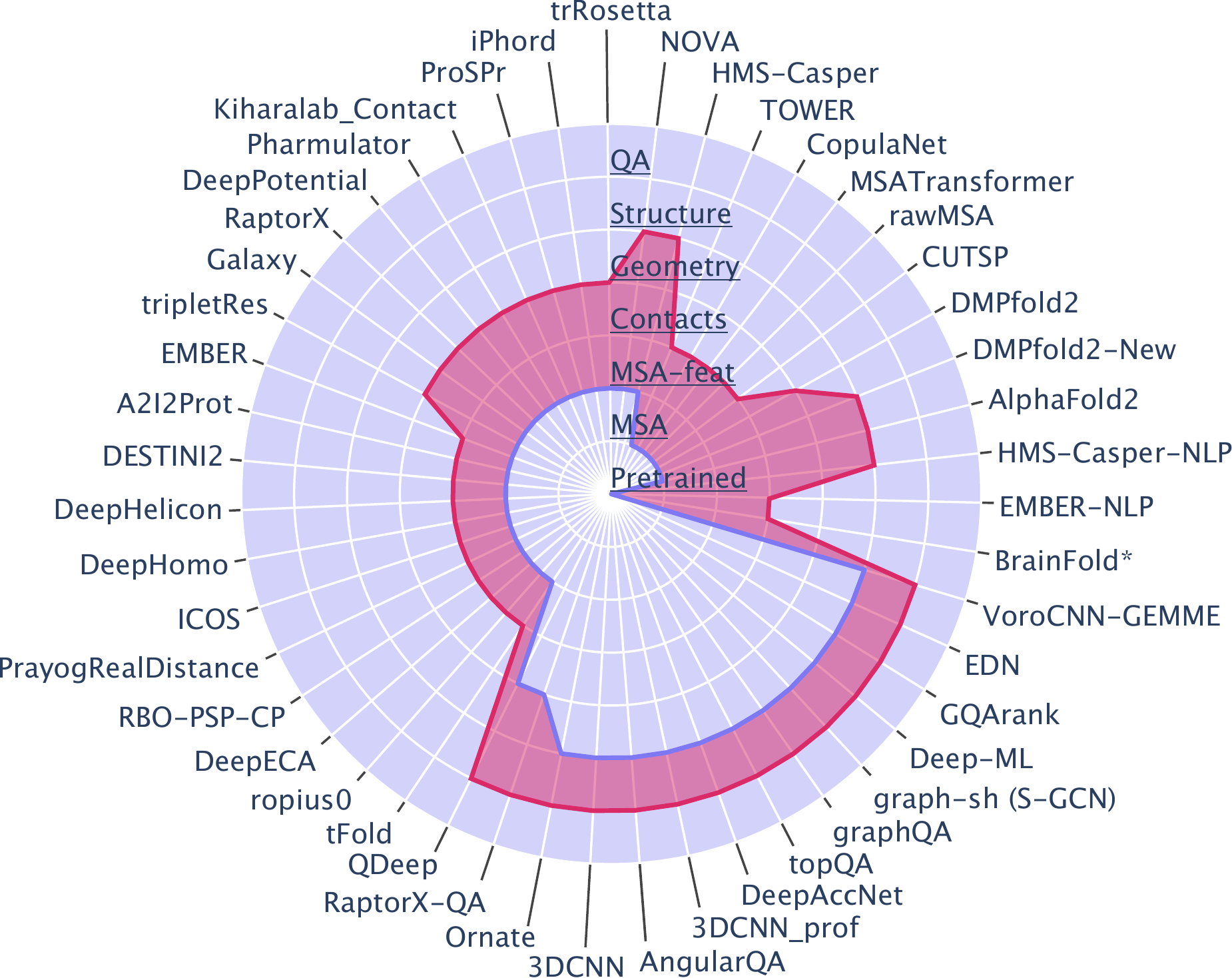}
\caption{\textbf{ 
Schematic representation of the inputs and outputs of deep learning-based methods in CASP14, excluding pipelines compiling several methods coming from different sources, and methods lacking a clear description.} The blue and red lines indicate the input and output levels, respectively. Pretrained: sequence embeddings determined from NLP models pre-trained on huge amounts of sequence data. MSA: raw multiple sequence alignement. MSA-feat: MSA features (such as PSSMs, covariance and precision matrices). Contacts: contact or distance matrix. Geometry: geometrical features, typically including contacts/distances and torsion angles. Structure: 3D coordinates. QA: model quality. In case of several inputs and/or outputs, we report those closest to the "end". BrainFold is highlighted with a star as it takes only the query sequence as input, without using pre-trained embeddings. This classification is based on available information from CASP abstracts and publications/preprints. See Supplementary Table S1 for more details.   }
\label{fig:overview}
\end{figure}

\section{End-to-end learning for protein structure prediction}

One of the advantages of deep learning methods compared with
traditional machine learning approaches is the ability to automatically extract features from
the input data without the need to carefully handcraft them (and potentially miss salient information). 
Assuming sufficient training data is available, learned features are expected to better generalize to heterogeneous or novel datasets. 
In
addition, it is generally accepted that  {\em end-to-end learning}, where the network is trained to produce the exact
desired output and not some sort of heuristic representation of it,
is advantageous. \replaced{Indeed, achieving a high accuracy on some intermediate result does not guarantee high accuracy on the final output. For instance, a learning algorithm may achieve a small loss on dihedral angles, and yet computing atomic coordinates from the predicted dihedral angles could lead to a high reconstruction error \cite{Elofsson:1995}.  Nevertheless, introducing well-chosen intermediate losses in a so-called "end-to-end" architecture can help to produce better final outputs \cite{jumper2021highly}. These auxiliary intermediate losses provide some guarantee that the method is not only able to produce an accurate final output ({\it e.g.} a protein 3D structure) but also to accurately model some other properties of the object under study ({\it e.g.} secondary structure, stereo-chemical quality...), and a mean to incorporate additional domain knowledge.}{An obvious benefit we may think of, is that when backpropagating the gradient all the way to the input, one is trying
to find a global solution to the problem. 
By contrast,
producing intermediate results would be equivalent to finding solutions to parts of the problem, which may not guarantee the global optimal solution.} While most protein structure prediction methods take pre-computed features as input and output a contact or distance map, possibly augmented with other geometrical features ({\bf Fig. \ref{fig:overview}}, see iPhord, ProSPr \cite{billingsprospr}, Kiharalab\_Contact \cite{subramaniya2020protein}, Pharmulator, DeepPotential, RaptorX \cite{xu2020improved}, Galaxy, TripletRes \cite{li2021deducing}, A2I2Prot, DESTINI2 \cite{gao2019destini}, DeepHelicon \cite{sun2020deephelicon}, DeepHomo \cite{yan2020accurate}, ICOS, PrayogRealDistance \cite{adhikari2020fully,adhikari2020deepcon}, RBO-PSP-CP  \cite{stahl2017epsilon}, DeepECA, ropius0 \cite{margelevivcius2020comer2}, tFOLD, plus QUARK, Risoluto, Multicom \cite{wu2021deepdist} and those from the Zhang lab),
 several efforts have been recently engaged towards developing \replaced{end-to-end}{so-called "end-to-end"} architectures. 
 Here, we will shortly review these efforts and try to identify the key components of what represents end-to-end learning in protein structure prediction ({\bf Table \ref{tab:end}}).

Ideally, the ultimate input would be the sequence of the query protein. \added{So far, only a couple of learning methods have exploited solely and directly this information to efficiently fold proteins {\it de novo} \cite{jumper2018trajectory,ingraham2018learning}. They rely on differentiable \cite{jumper2018trajectory} and neural \cite{ingraham2018learning} potentials whose parameters are learnt from conformational ensembles generated by Langevin dynamics simulations.} \replaced{More commonly, the strategy of state-of-the-art methods}{However, we are not aware of any state-of-the-art method relying only on this information. The common strategy} is to leverage the very high degenerative nature of the sequence-structure relationship through the use of a multiple sequence alignment (MSA) of evolutionary-related sequences, or a pre-trained protein language model (see below). 
In this context, methods qualifying for "end-to-X" learning should take as input raw (possibly aligned) sequence(s), as opposed to  
features derived from them such as conservation levels ({\it e.g.} stored in a Position-Specific Scoring Matrix or PSSM) or co-evolution estimates ({\it e.g.} mutual information, direct pairwise couplings).
One of the first examples
of end-to-X method was rawMSA \cite{Mirabello31415569}, which leveraged embedding techniques from the field of NLP, to map the amino acid residues into a continuous space adaptively learned based on the sequence context ({\bf Table \ref{tab:end}}). 
In DMPfold2 \cite{kandathil2020deep,kandathil2021ultrafast},
this idea was extended to MSAs of arbitrary lengths by scanning individual columns in the MSA with stacked Gated Recurrent Unit (GRU) layers. \added{CopulaNet \cite{ju2021copulanet} adopts a query-centered view by expanding the input MSA to a set of query-homolog pairwise alignments prior to embedding it. In AlphaFold2 \cite{jumper2021highly}, the MSA embedding is obtained through several rounds of self-attention (see below) applied to the MSA rows and columns. Beyond computing MSA embeddings, rawMSA, CopulaNet and AlphaFold2 add an explicit step aimed at converting the information they contain into residue-residue pairwise couplings through an {\em outer product} operation on the embedding vectors.} \added{Recently, a compromise end-to-X solution where the computation of traditional hand-crafted features takes place on the GPU and is tightly coupled to the network was implemented into trRosetta \cite{Yang31896580}, allowing for backpropagating gradients all the way to the input sequences \cite{norn2021protein}.}

At the other end of the spectrum, the ultimate output is the 3D structure of the query protein. Thus, an "X-to-end" deep learning architecture should directly produce 3D coordinates and not some intermediate representation such as a contact map.  
M. \citet{Alquraishi_2019_CellSystems} was among the first to develop such a method in
2019 ({\bf Table \ref{tab:end}}). 
The model takes as input a PSSM, without accounting for any co-evolutionary information, and
outputs the Cartesian coordinates of the protein. The torsion angles are predicted 
and used to reconstruct the 3D structure.
Although novel, such an approach has so far not proven to perform better than earlier
methods in CASP. One well-known problem is that internal coordinates are extremely sensitive to small deviations as the latter easily propagate through the protein, generating large errors in the reconstructed structure~\cite{Elofsson:1995}. 
To overcome this problem, it is possible to efficiently reconstruct Cartesian coordinates from a distance matrix by using multi-dimensional scaling (MDS) or other optimization techniques as
in CUTSP \cite{drori2019accurate}, DMPfold2 \cite{kandathil2020deep}, or E2E and FALCON-geom methods of CASP14. \added{In its classical formulation, used by both DMPfold2 and E2E, MDS extracts exact 3D coordinates (provided that the distance matrix is exact) through eigendecomposition of the centered distance matrix.
Nevertheless, one issue with using MDS as the final layer in the network is that the output may be a mirror image (chiral version) of the protein. The most recent version of DMPfold2 (DPMfold2-new in Table \ref{tab:end} \cite{kandathil2021ultrafast}) attempted to resolve this issue by adding an extra-GRU layer.} 
\added{AlphaFold2 takes a different route and elegantly solves the 3D reconstruction and the mirror-image problems jointly by learning spatial transformations of the local reference frames of each of the protein residues. Computing the geometric loss function in the local frames automatically distinguishes the mirror images, as one of the local axes is a vector product of the two others.}
Noticeably, even though X-to-end approaches generate a 3D structure, the latter is usually {\em refined} afterwards (for example through molecular dynamics simulations). \added{For instance, relaxation of AlphaFold2's output with a physical force field is necessary to enforce peptide bond geometry \cite{jumper2021highly}.}

Although the protein 3D structure appears as an obvious and legitimate target, one may wonder whether generating 3D coordinates confers any advantage, in terms of problem solving and performance, compared to a \emph{perfect} 2D contact map. First, as mentioned above, efficient methods to use 2D information for generating 3D models exist \cite{Adhikari25974172,Yang31896580}. Further, the most popular residue- or even atom-level loss functions used in deep neural networks (DNNs) do not depend on the superposition of the predicted model to the ground-truth structure
and are evaluated using the comparison of distance maps.  The most illustrative example is the local distance difference test (lDDT) \cite{LDDT}, which has been
employed as a target function in CASP14 by some of the best performers including AlphaFold2 \cite{jumper2021highly} and Rosetta. The value of this loss would not change if we swap the 3D and 2D representations. Nevertheless, it is not clear whether a perfect 2D map can be reached without using some 3D knowledge about the structure. Operating on 3D representations 
allows calculating global or local quality scores reflective of the structural accuracy in a way that 2D distance maps do not, \added{as illustrated by the mirror-image issue mentioned above}. The DNN can then learn to regress against these quality scores, and iteratively refine a first rough 3D guess by predicting (local) deformations to arrive at a better structure. \replaced{However, operating in 3D poses specific challenges related to the preservation of symmetries, which we discuss in Section 5.}{Nonetheless, operating in 3D poses some specific challenges linked to {\em equivariance}, which we discuss below.}
So far, the only successful
example of indisputable improvement of 3D structure representation over 2D maps is given by AlphaFold2 \cite{jumper2021highly}. Whether similar performance can be achieved with 2D maps and whether 2D maps are needed at all in the predictive process remain open questions.

Being able to produce 3D models resembling experimental structures implies being able to tell apart "good" from "bad" models. 
Hence, protein model quality assessment (MQA or QA), now referred in CASP to as estimation of model accuracy (EMA), has always been an important step in protein structure prediction pipelines. 
It allows, in principle, to choose the best models (in case of {\em global} QA) and/or spot inaccuracies in the proposed models for a subsequent refinement (in case of {\em local} QA). 
In recent years, a large number of deep learning-based approaches have been specifically designed for this task. 
Classically, they take a 3D model as input and then assess its quality in a stand-alone fashion ({\bf Fig. \ref{fig:overview}}). 
Alternatively, some teams proposed integrative approaches. 
For example, QDeep QA predictions \cite{shuvo2020qdeep} are based on distance estimations from DMPfold \cite{Greener31484923}. 
In GalaxyRefine2 \cite{lee2019galaxyrefine2}, RefineD \cite{bhattacharya2019refined}, and Baker suite \cite{hiranuma2021improved}, the QA is incorporated into a model refinement pipeline. 
Finally, QA blocks may be used as an integral part of a sequence-to-structure prediction 
process, as is the case in DMPfold2 \cite{kandathil2020deep} and AlphaFold2 \cite{jumper2021highly}.

\begin{table*}[bt]
\caption{Overview of X-to-end and end-to-X deep learning approaches for protein structure prediction.}
\begin{threeparttable}
\begin{tabular}{p{2.7cm}p{10.6cm}}
\headrow
\multicolumn{2}{c}{\bf{End-to-end learning}}  \\
\showrowcolors
AlphaFold2\cite{jumper2021highly} & The MSA, along with templates, is fed into a translation and rotation equivairant transformer architecture, which outputs a 3D structural model \\

DMPfold2 (new)\cite{kandathil2020deep,kandathil2021ultrafast} & The MSA, along with the precision matrix, is fed into a GRU, which outputs a 3D structure\\

\headrow
\multicolumn{2}{c}{\bf{End-to-X learning}}  \\
MSA Transformer\cite{rao2021msa} & Transformer architecture \\

rawMSA\cite{Mirabello31415569} & The MSA is fed into a 2DCNN (the first convolutional layer creates an embedding) which outputs a contact map\\

CopulaNet\cite{ju2021copulanet} & Extracts all sequence pairs from the MSA and feeds them to a dilated resCNN  \\

TOWER & The network is trained with a deep dilated resCNN to predict inter-residue distances directly from the raw MSA \\

\added{trRosetta\cite{Yang31896580}} & \added{Computes traditional MSA features on the fly and passes them to dilated convolutional layers }\\

\headrow
\multicolumn{2}{c}{\bf{X-to-end learning}}  \\

NOVA\cite{wang2019improved} & Adopts DeepFragLib from the same team which uses Long Short Term Memory units (LSTMs), to output a 3D structure \\

DMPfold2\cite{kandathil2020deep} & The MSA, along with the precision matrix, is fed into a GRU, which outputs distances and angles (version used in CASP14)\\

HMS-Casper\cite{Alquraishi_2019_CellSystems} & Raw sequences plus PSSMs are given to a "Recurrent Geometrical Network" comprising LSTM  and geometric units and outputting a 3D structure \\

\hline  
\end{tabular}
\end{threeparttable}
\label{tab:end}
\end{table*}

\section{The importance of data and data representations}

The success of deep-learning methods is heavily grounded in the availability of large amounts of data, and the development of suitable representations structuring and expressing the information they contain.  The advent of high throughput sequencing technologies has widened the gap between the number of known protein sequences and known protein structures.  Genomics has become pre-eminent in terms of data scale, with an exponential growth \cite{stephens2015big,navarro2019genomics}. These huge amounts of data offer unprecedented opportunities to develop high-capacity models detecting co-variation patterns and learning the "protein language".

\subsection{Leveraging (meta-)genomics}
In the last few years, the accessible resources for unannotated sequences coming from {\em metagenomics} experiments have multiplied. They include
databases like NCBI GenBank \cite{benson2018genbank}, Metaclust \cite{steinegger2018clustering}, BFD \cite{steinegger2019protein},  MetaEuk \cite{levy2020metaeuk}, EBI MGnify \cite{mitchell2020mgnify}, and IMG/M \cite{chen2019img}. \replaced{Since CASP12}{In CASP14}, several teams attempted to exploit this type of data, mostly to increase the depth of the MSAs and obtain a more accurate estimation of (co-)evolutionary features. For example, RaptorX  \cite{xu2020improved},  methods from the Yang and Baker teams \cite{Ovchinnikov28104891,wu2020protein},  Multicom \cite{wu2021deepdist}, and GALAXY exploited metagenome data for contact prediction and distance estimation between residue pairs in combination with residual convolutional neural networks (resCNNs).
The HMS-Casper \cite{Alquraishi_2019_CellSystems,alquraishi2019proteinnet}, DMPfold2 \cite{kandathil2020deep} and AlphaFold2 methods \cite{jumper2021highly} exploited them directly to predict 3D structures. Regarding QA, DeepPotential from the Zhang lab and QDeep \cite{shuvo2020qdeep} leverage generated MSA profiles from metagenome databases.
To gather large amounts of sequences, coming from different sources, many teams relied on the DeepMSA algorithm \cite{zhang2020deepmsa}.
Most of the time, the sequences were integrated altogether in a single MSA. However, some methods proposed to combine several MSAs with different weights ({\it e.g.} Kihara's lab) or to select a few of them with high depth and/or variability ({\it e.g.} DeepPotential).
Noteworthily, deep learning is not only used to exploit sequence alignments, but also to generate them. 
For instance, the SAdLSA algorithm improves the quality of low-sequence identity alignments by learning the "protein folding code" from structural alignments \cite{gao2020novel}. NDThreader \cite{wu2020deep} and ProALIGN \cite{kong2020proalign} are specifically designed to optimally align the query with the template in template-based modeling. Both methods exploit predicted or observed inter-residue distances to improve the sequence alignments, a strategy that proved powerful already in CASP13 \cite{Ovchinnikov28104891,buchan2017eigenthreader,zheng2019detecting}.

\subsection{From MSA to query-specific embeddings}

The most traditional way to extract information from an MSA is to compute a probabilistic profile or a PSSM reflecting the abundance of each amino acid at each position. This type of representation has been very popular from the very first CASP editions.
Over the past 10 years, direct coupling analysis (DCA)-based models\added{\cite{Marks22163331}}, including Potts model and pseudolikelihood maximization \cite{Balakrishnan21268112,Kamisetty2013,ekeberg2014fast,seemayer2014ccmpred}, and Graphical lasso-based (low-rank) models \cite{friedman2008sparse, Jones22101153, ma2015protein} became widespread in the community. These statistical methods explicitly estimate residue pairwise couplings as proxies for 3D contacts. 
More recently, some meta-models \cite{Jones25431331,Skwark23658418},
correlation and precision matrix-based approaches \cite{li2019respre,fukuda2020deepeca,Yang31896580}, 
and a variety of of deep-learning models \cite{golkov2016protein,Wang_2017_PLoS,Jones29718112,Kandathil31298436,Greener31484923,gao2019destini,wu2020protein,adhikari2020deepcon,adhikari2020fully,li2021deducing,wu2021deepdist},
including generative adversarial networks for contact map generation and refinement \cite{yang2020gancon,subramaniya2020protein},
got widely used to capture the same type of co-evolutionary information. One limitation of these methods is that they estimate average properties over an ensemble of sequences representative of a protein family. Hence, they may miss information specifically relevant to the protein query. 
The DeepMind team circumvented this limitation with AlphaFold2 by \added{computing embeddings for} residue-residue relationships within the query and sequence-residue relationships between the sequences in the MSA and the query\added{, and making the information flow between these two representations.}
Alternatively, one may {\em transfer} the knowledge acquired on hundreds of millions of natural sequences to generate query-specific embeddings ({\bf Table \ref{tab:nlp}}). Several models developed for NLP, including BERT \cite{devlin2018bert}, ELMo \cite{peters2018deep}, and GPT-2 \cite{radford2019language}, have been adapted to the "protein language". During the semi-supervised training phase, the model attempts to predict a masked or the next token \cite{rao2019evaluating}. In CASP14, EMBER directly made use of ELMo and BERT while HMS-Casper \cite{Alquraishi_2019_CellSystems} used a reformulated version of the latter, called AminoBert. A2I2Prot and CUTSP leveraged the TAPE initiative \cite{rao2019evaluating}, which provides data, tasks and benchmarks to facilitate the evaluation of protein transfer learning. 

\subsection{Representations of protein structure}

\begin{figure*}[bt]
\centering\includegraphics[width=\textwidth]{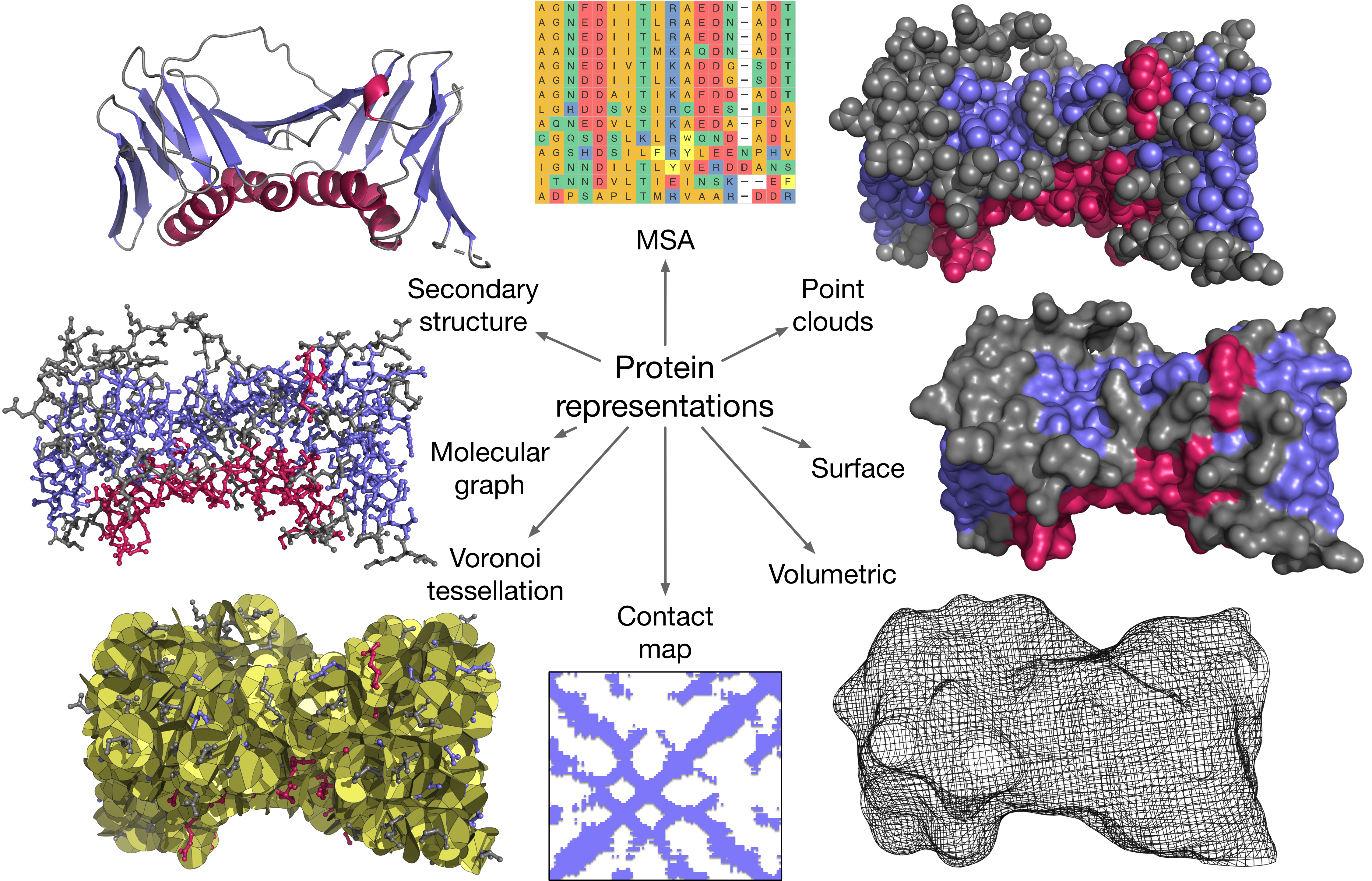}
\caption{{\bf Comparison between protein representations for human PCNA (PDB code:1AXC, chain A) .} 
}
\label{fig:representations}
\end{figure*}

Sequence-based protein representation may be enriched with different levels of {\em structural information}, for example, some prior knowledge about secondary structure (SS) elements.
In principle, some of these elements, such as alpha helices or beta strands, can be represented with 3D primitives. An interesting idea that we saw in CASP14 was the use of a discrete version of Frenet-Serret frames for the protein backbone parametrization by HMS-Casper.
However, such a representation is very complex, and a much simpler way would be to abstract SS primitives with a hydrogen-bond (HB) 2D map.
For example, the ISSEC network was specifically trained to segment SS elements in 2D contact maps \cite{zhang2020issec}.
Similarly, the protein 3D topology may be abstracted as a 2D contact map, or its probabilistic generalization, {\it e.g.} a matrix filled with continuous probabilities or contact propensities
between protein atoms or residues. Beyond 2D contact maps, richer descriptions of the 3D structures can be achieved with 2D contact manifolds and protein surfaces, 3D molecular graphs, point clouds,
sets of oriented local frames, 
volumetric 3D maps, or
3D tessellations, {\it e.g.} through Voronoi diagrams ({\bf Table \ref{fig:representations}}). These different levels of protein representations and their applications in CASP are discussed in more details below
and schematically shown in {\bf Fig. \ref{fig:representations}}.

\begin{table*}[bt]
\caption{Overview of approaches transferring knowledge from large amounts of protein sequence data.}
\begin{threeparttable}
\begin{tabular}{p{2.9cm}p{10.4cm}}
 \hline
HMS-Casper (NLP)\cite{Alquraishi_2019_CellSystems} & Sequence embeddings generated by a reformulated version of the BERT language model are given as input to a LSTM-based architecture \\
    EMBER (NLP) & Sequence embeddings are generated by BERT and ELMo trained
    on protein sequence sets and given to a resCNN with dilatations\\
    
    A2I2Prot & A sequence embedding correlation map is fed into a resCNN\\
    
    CUTSP\cite{drori2019accurate} &  Sequence embeddings, along with a MSA, are fed to a bi-directional GRU and LSTM with skip connections, followed by an Encoder-Decoder architecture\\

\hline  
\end{tabular}
\end{threeparttable}
\label{tab:nlp}
\end{table*}

\subsubsection {Volumetric protein representations}

The first attempt to train 3D CNNs on a volumetric protein representation dates back to CASP12, with the goal of assessing protein model quality  \cite{Derevyanko29931128}.
The architecture was robust but had two major limitations.
Specifically, it relied on a predefined protein's atom types, and the orientation of the protein model given as input had an influence on the output of the network. In other words, the network was not {\em rotation-invariant}. 
To cope with this issue, it had to be trained on the input data augmented by a set of rotations applied to each input protein model. In a follow-up work, \citet{derevyanko2019protein} introduced an  SE(3)-invariant architecture building on \citet{Weiler2018}.

The Ornate architecture overcomes both limitations \cite{Pages30874723} ({\bf Table \ref{tab:qa}}).
Ornate learns atom type embeddings and constructs {\em local} volumetric representations of each amino acid in a protein in a local coordinate system,
thus achieving {\em local translation-rotation invariance} of the network.
Sato-3DCNN by \citet{sato2019protein} used an idea similar to that of Ornate with oriented local frames but did not automatically learn the atom type embeddings.
3DCNN\_prof (or P3CMQA) extended this network with additional input features including MSA profile, predicted secondary structure, and solvent accessibility \cite{takei2021p3cmqa}.
Finally, DeepAccNet showed remarkable performance in the CASP14 refinement category. This architecture extends Ornate
by adding 1D  and 2D input features coming from sequence and Rosetta energy terms  \cite{hiranuma2021improved} ({\bf Table \ref{tab:qa}}). 
It predicts per-residue model accuracy and also inter-residue distance signed error, such that
the network can be efficiently used for protein model refinement.


\subsubsection {Graph protein representations}
A remarkable fact of the CASP14 edition is the emergence of graph representations as means to encode sequence and/or structure information. 
Indeed, graphs allow formally and compactly encoding diverse relationships between heterogeneous objects. 
The DeepMind team was probably the one who best exploited this property, by using the graph representation to encode both sequence information taken as input and structural information learned by the architecture in an end-to-end fashion. Several other teams have made contributions toward deriving graph representations for protein data and developing algorithms operating on these graphs. For example, DeepML, GQArank, LAW, and GraphQA \cite{Baldassarre32780838} applied classical graph convolutional networks (GCNs) at the residue level, where the convolution operator averages the features of each node's neighbours ({\bf Table \ref{tab:qa}}).
Spherical Graph Convolutional Network (S-GCN) made a step further and 
extended the graph convolution operator for spherical geometry in molecular graphs. 
This allowed to effectively encode mutual angular dependence of neighbouring graph nodes using spherical harmonics expansions
\cite{igashov2020spherical}. 
In its turn, GNNRefine predicted distances between protein atoms using graph neural networks
and then converted  these distances into interatomic potentials and employed them for protein structure refinement \cite{jing2020fast}.
A more recent method, GVP-GNN \cite{jing2021learning}, augments graph networks with the ability to reason about protein features expressed as geometric vectors in an equivariant manner. 


\subsubsection {\deleted{From graphs to}Point clouds \added{and oriented frames}}
\deleted{In principle, a A graph can has two extreme forms,  {\em complete}, with every pair of nodes connected by an edge, or {\em empty}, without any edge.
these two forms can be particularly useful for protein topology description.
If it is not known it might be better to use the complete graph description.}

Alternatively \added{to a graph}, one may want to make the protein topology evolve through the architecture, without explicitly fixing it. The protein is then seen as a set of isolated nodes with specific positions in 3D space, in other words, a {\em 3D point cloud}.  
The EDN method in CASP14 was the first approach to describe a 3D protein structure as a set of points in 3D \cite{eismann2020hierarchical,eismann2020protein}. In this setting, each point is associated with a set of features. At input, individual\deleted{s} atoms are represented as points and the chemical element type is the sole associated feature. New point-based features are then calculated over a series of rotation-equivariant convolutions based on the 3D environment around each point. In addition, the network aggregates information at different levels of point hierarchy, from individual atoms over $\alpha$-carbons to the whole protein.

\added{In AlphaFold2, the protein structure is modelled through a related representation, a cloud of local coordinate system frames, each unambiguously associated with a residue. The method updates these frames indirectly by applying an attention mechanism to "3D points" generated from the query sequence embedding. Hence, at each step, the changes computed on the embedding, which is an abstract representation of the protein, are converted into concrete 3D coordinate changes. DeepMind team uses the term "residue gas" to refer to this frame cloud representation, to emphasize the fact that the peptide bond geometry is totally unconstrained ({\it i.e.} the peptide bonds are "broken"). }

\subsubsection {3D tessellations}

A somewhat similar concept to the graph description is a tessellation of the 2D or 3D space. Tessellations are partitions of the space (3D for most CASP applications) into regions (cells) with specific properties. A tessellation can be represented with a graph, where each node stands for a cell and each edge for the contact between two cells.
A particularly useful type of tessellation is the {\em Voronoi tessellation}, or {\em Voronoi diagram}. Considering protein structure, the interior of the Voronoi cell around each protein atom must be closer to that atom than to any other.
As the atoms are physical objects with different radii, the Voronoi cells are defined by intersecting pairwise bisector surfaces. 
In case of the additively-weighted Voronoi tessellation, a bisector surface is a part of a hyperboloid of two sheets, approaching a plane when the difference between atomic radii tend to zero ({\bf Fig. \ref{fig:representations}}).
The Voronoi tessellation turned out to be a very powerful description of protein structure and interactions and has been used in structural bioinformatics for several decades \cite{pontius1996deviations,zimmer1998new,poupon2004voronoi,cazals2006revisiting,olechnovivc2014voronota}.
In CASP14, we saw this description incorporated into DNNs.
VoroCNN was the first attempt to construct a deep network passing messages between the neighbouring Voronoi cells \cite{igashov2020vorocnn}. The network performs a hierarchical tessellation by starting at the atom level, and then aggregating features to the residue level. Another interesting idea was implemented in VoroMQA-dark \cite{Dapkunas:2021tg}, an extension of the VoroMQA method \cite{Olechnovic28263393} where contact Voronoi areas and pseudo-energies are fed to a feed-forward network.
An important particularity of VoroMQA,  VoroCNN, and related methods is that their Voronoi tessellations are constrained by the solvent-accessible surface.
Therefore, the Voronoi cells of the surface atoms are finite, and the corresponding contact surfaces  abstract solvent-protein interactions.

\subsubsection {2D manifolds}

Another flavour of the VoroMQA method has also proved successful in scoring protein complexes since the CASP12-CAPRI experiment \cite{dapkunas2019structural,olechnovivc2019voromqa}. 
Here, only the protein-protein contact areas contribute to the final score. In such an approach, the 3D protein structures are viewed as {\em 2D surface manifolds}. 
In CASP14, this type of protein description has been combined with deep learning \cite{Dapkunas:2021tg}. Other developments include the application of ideas from the recent MoNet manifold network architecture \cite{monti2017geometric} toward learning protein surfaces with very exciting outcomes for protein binding sites and protein-protein complexes prediction
\cite{gainza2020deciphering,sverrisson2020fast}. 
Overall, 2D surface  manifolds seem very powerful and compact representations of 3D shapes, at least in the context of protein-protein interactions. 
\begin{table*}[bt]
\caption{Overview of deep learning QA approaches in CASP14.}
\begin{threeparttable}
\begin{tabular}{p{3.8cm}p{9.5cm}}
\headrow
\multicolumn{2}{c}{\bf{Volumetric representations}}  \\
3DCNN\cite{Derevyanko29931128} & A non-invariant 3D CNN  \\
Ornate\cite{Pages30874723}& A local frame-based 3D CNN  model with learned atom embeddings \\
Sato-3DCNN\cite{sato2019protein}& A local frame-based 3D CNN \\
3DCNN\_prof (P3CMQA)\cite{takei2021p3cmqa}& Extends  Sato-3DCNN with predicted features and PSSMs \\
SE(3)-3DCNN\cite{derevyanko2019protein} &  An invariant 3D CNN based on \cite{Weiler2018} trained for protein complexes \\
iPhord \& DeepMUSICS & 3D CNNs \\
TopQA\cite{smith2020topqa} &  3D CNN with explicit rotations and automatic scaling to fit into a unit cube\\
DeepAccNet\cite{hiranuma2021improved} & Extends Ornate with 1D and 2D geometrical features to predict per-residue model accuracy and also inter-residue distance signed error\\
\headrow
\multicolumn{2}{c}{\bf{Graph representations}}  \\
 Graph-QA\cite{Baldassarre32780838} & GCN with representation learning,
explicit modeling of both sequential and 3D structure, geometric
invariance, and  computational efficiency\\
S-GCN\cite{igashov2020spherical} & Molecular-graph-based method where angular information is accounted for using spherical harmonics \\
  GQArank & GCN with many features, including PSSM and predicted geometrical properties \\
  DeepML  & A classical GCN \\
LAW & 5-layer GCN followed by a 3-layer 1D CNN\\
  \headrow
\multicolumn{2}{c}{\bf{Tessellations, 2D manifolds, and point clouds}}  \\
VoroCNN\cite{igashov2020vorocnn} & A CNN built on a hierarchical 3D Voronoi tessellation of a protein molecule\\
VoroMQA-dark\cite{Dapkunas:2021tg} & A CNN-based extension of VoroMQA \cite{Olechnovic28263393} \\
EDN\cite{eismann2020protein} & A point-cloud representation of the atomic structure combined with rotation-equivariant, hierarchical convolutions \\ 
\hline  
\end{tabular}
\end{threeparttable}
\label{tab:qa}
\end{table*}

\section{From convolutions to attention}

\begin{figure}[bt]
\centering\includegraphics[width=1\columnwidth]{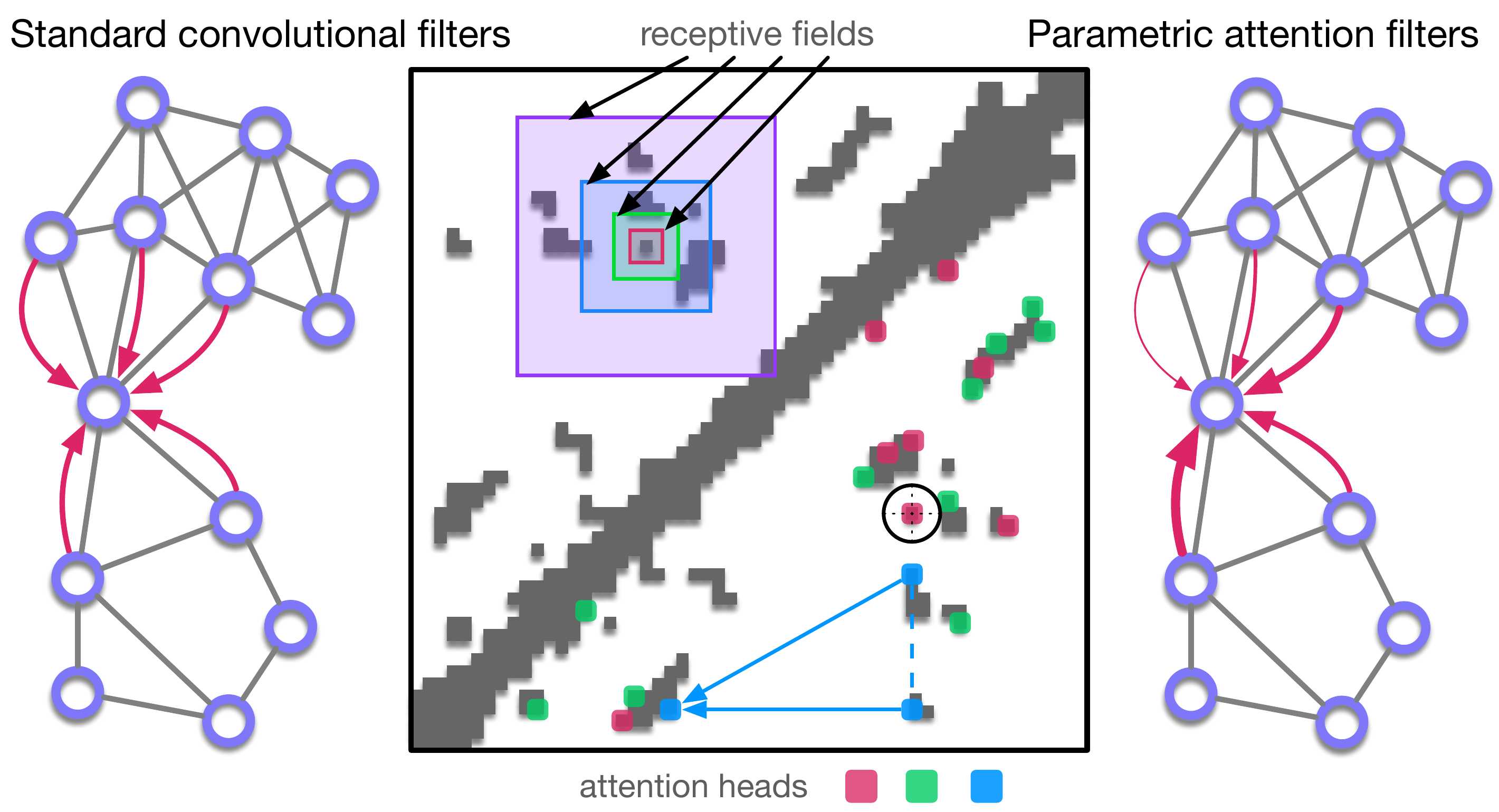}
\caption{{\bf Comparison between convolutional filters and attention  heads.} 
The input data is either represented as a 2D image or as a graph. The information encoded may be for instance MSA-inferred covariances or template-derived Euclidean distances between protein residues. In the top left triangle of the image, the overlapping squares correspond to the increasing receptive fields obtained by stacking multiple layers of convolutional filters in a 2D-CNN. In the area under the main diagonal, the coloured squares represent the input points being the most important with respect to the one in the center of the circle, as the result of applying {\em attention filters}. In case the nodes represent residues, and the attention weights can be interpreted in terms of 3D distances or contact, only 2 links are necessary to infer a triangle (in blue). \added{This observation was exploited by AlphaFold2 through the implementation of {\em triangular self-attention} \cite{jumper2021highly}. } In the graph on the left, the red arrows indicate a standard convolution aggregating information from neighbouring nodes. On the right, the attention mechanism puts more weight on certain neighbours (illustrated by arrow thickness).}
\label{fig:attention}
\end{figure}

The choice of the protein data representation is intimately linked to that of the deep learning architecture and operators. Historically, the first deep learning breakthrough in protein structure prediction came from CNNs, widely used for computer vision, applying multiple filters to protein "images". Each filter of a standard CNN aggregates information coming from a region of the input data, namely the receptive field (\textbf{Fig. \ref{fig:attention}}, area above the main diagonal). The filters in the first layer directly operate on the input data, while the filters in each of the subsequent layers apply some operation on the output of the previous layer. As the information is processed by the successive layers, the size of the receptive field increases, and, as a consequence, longer-range dependencies are captured. However, this accounting of long-range dependencies comes at the expense of precision, since it occurs only after a certain depth in the network. 
Indeed, the late layers corresponding to a large receptive field do not directly operate on the input data but on some abstract representation of it containing less information. 
This makes CNNs strongly dependent on the way the input observations are ordered or located with respect to each other. For example, when dealing with a 2D covariance matrix computed from a multiple sequence alignment, local patterns formed by residues adjacent in sequence  
will be captured with higher precision. 
This may constitute a limitation since protein 3D structures are also stabilized by interactions formed between distant amino acids in the sequence. 
When treating a raw MSA as a 2D image, the order of the sequences will also have an influence while this order may somewhat be arbitrary. 
In the case of a geometric representation of the 3D protein structure, the information encoded in local neighbourhoods of the Euclidian space will be aggregated first and thus more precisely captured than relationships between distant atoms.

One way to overcome such limitation is to introduce gaps (dilations) 
when defining the filters. With a dilation $d$, the window starting at
location $i$ of size $k$ is
$[x_i\;x_{i+d}\;x_{i+2d}\;\cdots\;x_{i+(k-1) \dot d}]$. Stacking
dilated convolutions with increasingly large $d$ allows operating on
exponentially large receptive fields, while retaining short
backpropagations~\cite{yu2016multiscale,gupta2017dilated,Senior31942072}.
In CASP14, {\em dilated convolutions} were used by several groups, including 
    ProSPr  \cite{billingsprospr}, DESTINI2  \cite{gao2019destini}, CopulaNet \cite{ju2021copulanet}, 
    PrayogRealDistance  \cite{adhikari2020fully,adhikari2020deepcon},
    and also EMBER, TOWER, ICOS, and LAW/MASS. 
Another solution lies in the
{\em self-attention mechanism}, where parametric filters capture high-order
dependencies between the input observations at arbitrary range and
with high precision (\textbf{Fig. \ref{fig:attention}}, area under the main diagonal). The intuition is to focus on the most relevant
parts of the input with respect to a task or output (general
attention) or to another part of the input
(self-attention). Specifically, for each input point, a set of
trainable attention weights determines the relative importance of each
of the other input points. Attention mechanisms have made a major
breakthrough in NLP, as they allow
keeping in memory the sequence context (although limited in practical applications) in translation tasks~\cite{bahdanau2016neural,gehring2017convolutional,vaswani2017attention,choromanski2020rethinking}.
They are particularly well suited \replaced{for}{to} data whose underlying representation
does not have a grid-like structure and rather lie in an irregular
domain. Such data can often be represented in the form of
{\em graphs}. While standard graph convolutions indifferently aggregate
information from neighbouring nodes \cite{kipf2017semisupervised} (\textbf{Fig. \ref{fig:attention}}, left graph), the
attention mechanism puts more importance on a subset of neighbours, without increasing the time complexity \cite{velickovic2018graph} (\textbf{Fig. \ref{fig:attention}}, right graph). 
In the extreme case, each node may attend to all
other nodes, allowing for a full inference of the graph
structure. 
This strategy \replaced{has}{may have} been employed by the DeepMind team in CASP14
\added{ in several places of their AlphaFold2 architecture. 
Particularly, when operating on the 3D structure, they impose a strong spatial/locality bias on the attention that depends on the relative positions of 3D points produced in the local frames of the residues (invariant point attention, IPA).
}
\deleted{Very }Recent works have also shown that the residue-residue dependencies extracted by certain attention heads in transformers trained on large amounts of sequences can be directly interpreted as 3D contacts or distances \cite{bhattacharya2020single,rao2020transformer,rao2021msa}.

\section{From invariance to equivariance}

\begin{figure*}[h]
\centering
\includegraphics[width=1.0\textwidth]{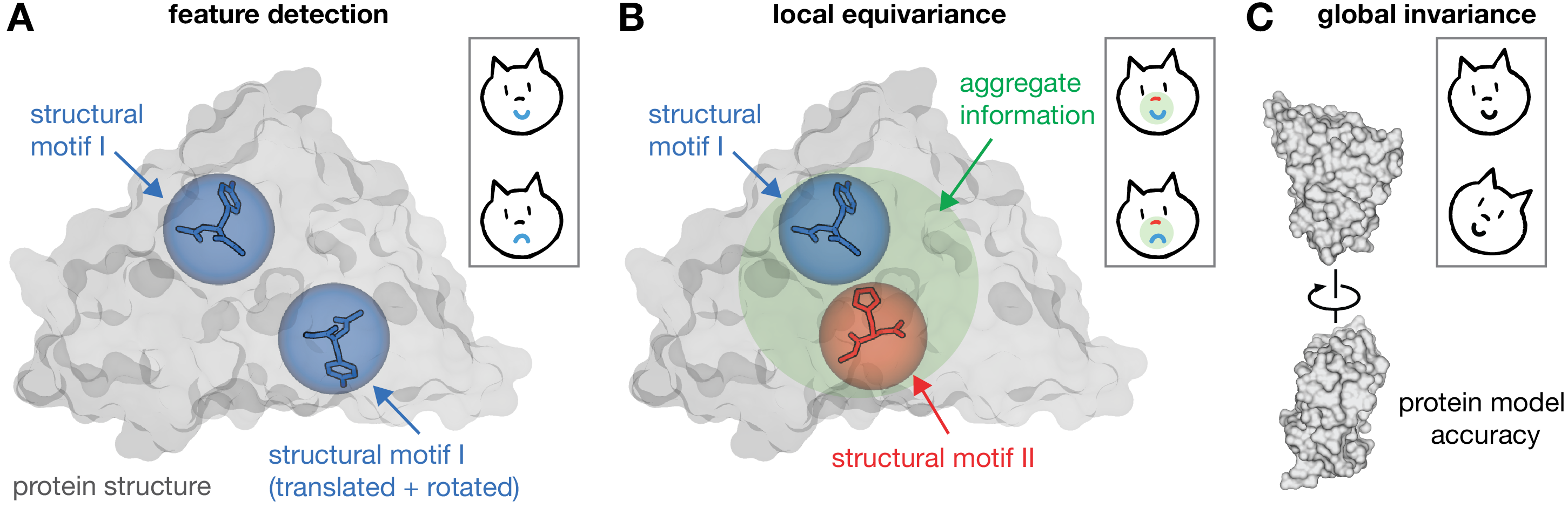}
\caption{\textbf{Symmetry considerations in learning from macromolecular structure} \textbf{(A)} Given the 3D structure of a protein (grey surface), it is desirable that a neural network can identify a structural motif (blue) --- a specific arrangement of atoms in 3D --- independent of the position and orientation at which the motif occurs in the structure.  Insert: Analogy with a cat face, where the 'structural motif' is represented by the mouth of the cat.
\textbf{(B)} A larger receptive field encompassing two motifs (blue and red) is shown in green. In order to aggregate information from local neighbourhoods, independent motif detection is not enough. The network requires information on the relative orientation and position of the structural motifs, which is realised through translation and rotation equivariant features. Insert: The cat's happiness changes with the relative orientation of the nose and mouth `motifs`. \textbf{(C)} At the global level, the accuracy of a protein model is rotation invariant. Insert: A global rotation leaves the cat's happiness unaffected.}
\label{fig:equivariance}
\end{figure*}

Breakthrough applications of deep learning often have in common that the underlying methods cater to specific characteristics of the data domain, such as long-range dependencies in text and hierarchical features in images. Figure~\ref{fig:equivariance} considers relevant characteristics for macromolecular structure, namely invariance and equivariance with respect to translations and rotations in 3D. For illustration purposes,  the figure includes a series of cat cartoons in 2D.  

As training progresses, a neural network should learn to identify structural features helpful for the task it is designed to solve. We refer to such an informative feature as a \textit{structural motif} --- a specific arrangement of a set of atoms in 3D (Figure~\ref{fig:equivariance}A). In the case of the cat cartoons, the mouth and nose of the cat shall correspond to “structural motifs”.  Given a protein structure, a network should further be able to identify structural motifs independent of the orientation and position in which they occur. 
If this ability is not built into the network architecture, the network needs to learn it by seeing the same motif in different orientations and positions. This additional learning task does not only require more network parameters, but the network can also only learn an approximation of the desired detection ability. 

The general idea is that incorporating specific domain knowledge into the architecture --- here the assumption that a structural motif is the same independent of where and in which orientation it occurs --- provides an advantage over a more flexible network architecture through a reduction in model parameters that ultimately translates to better network prediction accuracy given the same, finite amount of training data. Assuming the domain-specific assumptions are true, this reduction in model parameters does not result in a loss of expressive power, as it only prevents the network from learning functions inconsistent with the assumptions \cite{bronstein2021geometric}.

Feature detection independent of orientation and position is not enough 
in the case of a larger receptive field encompassing two structural motifs (Figure~\ref{fig:equivariance}B). To aggregate information from local neighborhoods, the network also needs information on the relative orientation and position of the learned structural motifs. These geometrical aspects are crucial since they govern the intramolecular interactions. The importance of relative orientation is also apparent in the cat cartoons --- rotating the mouth motif by 180$^\circ$ with respect to the nose turns the happy cat into a sad one. Here, the desired property is that of {\em equivariance}.
Informally, a function or neural network layer is equivariant to some transformation (such as a rotation or
translation) if a transformation of the input results in the same transformation of the output. For {\em invariance}, the function output does not vary with respect to transformations of the input.
Standard CNNs and other neural network architectures already account for translational equivariance ({\it e.g.}, by encoding only relative positions of atoms) but not for rotational equivariance. 
Rotational symmetry is specifically important in 3D: non-equivariant architectures require a factor of $\mathcal{O}(\delta^{-1})$ more filters to achieve an angular resolution of $\delta$ in 2D but already $\mathcal{O}(\delta^{-3})$ more filters in 3D \cite{thomas2018tensor}. 

Finally, invariance {\it vs.} equivariance can also depend on the perspective, as illustrated in Figure~\ref{fig:equivariance}C. 
If the goal is to predict global protein model accuracy, such as measured by global lDDT, a network should provide equivariant outputs at the local level but a global prediction that is invariant under rotations and translations. Turning again to the cartoon cat, we similarly note that a global rotation of the cat leaves its happiness unaffected. 
 
Historically, machine-learning-based scoring functions \cite{Wallner:ProQ,karasikov2019smooth} were inspired by statistical potentials \cite{miyazawa1985estimation,Sippl:1990} relying on pairwise distance/ angular distributions or contact maps, which are perfectly rotation and translation invariant representations. The challenge of equivariance arose with the development of deep-learning architectures operating directly on raw 3D geometry, rather than precomputed (primarily 1D or 2D) features \cite{Derevyanko29931128}. \citet{Pages30874723,sato2019protein} elegantly circumvented the need for rotational data augmentation in standard CNNs by leveraging a residue-level coordinate system to learn an invariant local quality metric. \added{AlphaFold2 also represents the protein residues with local oriented frames. Both the frames and their relationships are learnt through a geometry-aware attention mechanism.} Graph representations of protein structure generally similarly encode the local and global 3D geometry through rotation-invariant scalar features such as angles and distances \cite{Baldassarre32780838}. Recent efforts  include the use of spherical convolutions in combination with a residue-level coordinate system to learn a local quality metric \cite{igashov2020spherical}, and the development of invariant volumetric  \cite{derevyanko2019protein} and 
equivariant point clouds representations in 3D  \cite{eismann2020hierarchical, eismann2020protein}.
Specifically, in \citet{eismann2020protein},
starting from the 3D coordinates and element type of each atom, the network first learns rotation-equivariant representations of local neighborhoods and then aggregates this information hierarchically to predict a rotation-invariant fingerprint at the level of the entire protein structure, reflecting the previously discussed symmetry considerations. 
The architecture builds on tensor field networks \cite{thomas2018tensor} in which points in 3D space are associated with tensor features (such as scalars and vectors) and these features are updated over consecutive network layers. 

From a broader perspective, early work from
\citet{cohen2016steerable} pioneered the use of tools from group representation theory to build a rotation equivariant neural network architecture. This idea has been followed by a rich body of publications, including translation and rotation equivariant architectures for 3D point clouds \cite{thomas2018tensor, kondor2018n, anderson2019cormorant}. These architectures can be seen as equivariant extensions of neural-network-based radial and angular symmetry functions for molecular structure \cite{Behler2007, Schutt2017, Smith2017}. \citet{Weiler2018} further proposed a rotation-equivariant architecture for continuous data in 3D. All these equivariant architectures  share the use of spherical harmonics, a set of functions defined on the unit sphere that is intrinsically linked to 3D rotation equivariance. 
Spherical harmonics have played a prominent role in molecular surface representations for several decades \cite{max1988spherical,ritchie1999fast}  and are also at the heart of the classical fast multipole method \cite{greengard1987fast}.



We believe that equivariant architectures in learning from macromolecular structure will grow further in popularity due to their parameter-efficient expressive power and their ability to directly reason about, and also predict geometric quantities such as vectors. The recent work by \citet{fuchs2020se3transformers} on small molecules and the success of AlphaFold2 at CASP14 further suggest promise for network architectures combining equivariance and attention mechanisms. \added{This combination was also leveraged in the most recent RoseTTAFold network \cite{baek_accurate_2021}.}

\section{Conclusion and outlook}

The success of CASP14 methods in general, and AlphaFold2 in particular, leads to the awareness of the community 
that highly-accurate protein structure predictions can be obtained for
virtually all well-folded protein domains, but also that 
 the performance gap between AlphaFold2 and other methods is significant. 
 After CASP13, it took the community about 12
months to catch up with AlphaFold1. It is uncertain whether we can expect a similar duration, in part due to the combination of innovative approaches in AlphaFold2. \added{We should also not forget that the computational costs engaged by DeepMind to develop the method, tune its hyperparameters and train it are substantial.}
A rough estimate for the cost of training the network architecture using cloud resources exceeds \$20K.
This emphasises the need for a community-wide effort to catch up with AlphaFold2 and/or the design of shortcuts alleviating training costs.

\added{Recently, a first attempt to reproduce DeepMind's work was presented. The RoseTTAFold method \cite{baek_accurate_2021} is not as accurate as AlphaFold2, but still clearly better than all earlier methods. Further, it is rather cheap to train and use, providing hope that computational cost will not be a burden for large scale application of these methods.} 
Nonetheless, it is clear to us that the entire community ought to come together with an open mind to develop next generation deep learning-based tools for protein structure prediction. 
Such an effort
would not only have an impact on the field of structure prediction but
also on related fields through the innovation of novel deep learning methods. Further,
there are, as discussed below, still many challenges in computational structural biology
that are not (yet) solved.

\subsection{The impact of accurate models in structural biology and bioinformatics}

Accurate 3D structures provide valuable information about protein
biological functions. They can be used by themselves, and also as a starting point for further computational studies. For some studies, the accuracy of theoretical models has been sufficient, for others not. The improvements brought by CASP14 will, therefore, increase the number of suitable targets for  tasks requiring a very high-quality models, such as mutational effects prediction, ligand binding site identification, molecular dynamics simulations, drug discovery and
enzymatic reactions modeling, to list a few.  They also open avenues for a tight cross-talk between structure {\it in silico} prediction and experimental determination. Already in CASP14, models from AlphaFold2 were used to phase crystals and thereby to solve protein structures. If this can be extended and done systematically, there are probably hundreds of unsolved protein crystals that could benefit from high-quality models. The latter can also help in the initial steps of single-particle Cryo-EM reconstructions. 
However, the full extent of the impact of computational models on structural biology and other fields will likely depend on their ability to provide profound novel biological insights, that are generally accepted by the community.
When this will happen depends on how good the models are. One possible start here could be to examine what additional information was obtained by the experimental structures, phased by the AlphaFold2 models, over the models themselves. If predicted models are accurate enough, one major role for future structural biology might become to identify all the chemical compounds (proteins, ligands, lipids, co-factors) that interact and then use artificial intelligence methods to predict the structure of this ensemble.

\subsection{Learning the laws of physics?}

Most methodologies in computational structural biology build on physics' first principles to describe individual atoms and how they interact.
These laws are then used to model larger molecules such as proteins. 
One may wonder to what extent the emergent data-driven approaches that do not explicitly implement detailed physical descriptions of biomolecules are able to implicitly learn physics laws. For instance, end-to-end sequence-to-structure deep learning methods do not explicitly model water molecules, co-factors or partners. Yet, AlphaFold2 was able to determine the residue side-chain orientations competent for binding a zinc ion in the M23 peptidase  (target T1056) and also the bound conformation of a cell wall surface anchor protein forming homo-trimers (target T1080). These conformations make sense, from a physical point of view, only when the co-factor or the partner is present. From a data science point of view, however, if some proteins are always found in complex with co-factors or partners, then the machine will learn to associate the matching sequence contexts with bound conformations. In other words, it implicitly learns the physical contexts  compatible (at least in the experimentally data at hand) with a particular sequence context.
This ability may be further exploited to discover ligand- or partner-binding sites by analysing the geometry and physico-chemical properties of the predicted conformations. Indeed, the machine may not only be able to predict plausible bound conformations but also to identify the location of the "missing" co-factor(s), ligand(s) or partner(s).
Next, it can be asked --- is an accurate physical description necessary for other tasks such as binding free energy estimation and mutant stability prediction?
The main limitation might be the amount of training data
available to develop such methods, but we certainly do foresee many
attempts to transfer these ideas to other areas of computational structural biology.

\subsection{Protein disorder, flexibility and dynamics}

Beyond 3D structure, proteins' dynamical behaviour is important for their functions. Flexibility is
necessary for binding, enzymatic reactions, transport, and many more \cite{frauenfelder1991energy, henzler2007dynamic}. Many proteins adopt two or more stable conformations and the equilibrium between these states has a direct implication on their functioning. For instance, protein kinases, representing about 2\% of the human proteome, adopt two distinct forms, one inactive and the other active. These two states are clearly distinguishable and can be captured by X-ray crystallography. 
As an extreme version of flexibility, intrinsic disorder is commonly observed in eukaryotic proteins and plays
crucial roles in transient protein-protein interactions as well as in linkers between domains. Some  intrinsically
disordered regions (IDRs) form a stable structure upon binding to their partners but it is difficult to experimentally identify them. 

Co-evolutionary patterns extracted from related sequences have proven useful to predict some IDR bound forms~\cite{TothPetroczy27662088} and, in some cases, to untangle a protein's multiple functional states  \cite{morcos2019role,sfriso2016residues,morcos2013coevolutionary}.
However, systematically training DL models to predict protein flexibility, either as a probabilistic structural profile or as conformational {\em multi-modalities}, remains very challenging. 
Experimental measurements are very scarce and/or probe conformational states only indirectly. For instance, crystallographic temperature (B-) factors, although abundant, are not reliable proxies of internal molecular fluctuations. Indeed, at cryogenic temperatures, the main contribution to B-factors will be crystal lattice disorder. Another option is to use nuclear magnetic resonance (NMR) data as the ground truth for structure prediction architectures.
However, we have relatively few NMR structures (6K that contain at least one protein chain as of 2021), and even fewer collected raw NMR observations.
In principle, one can train deep models on NMR-inferred 3D reconstructions, often given as multiple models in the PDB, instead of the raw NMR data, but this most likely does not reflect the true flexibility of a model,
as it is also dependent on the number of nuclear Overhauser effect (NOE) constraints obtained in the NMR experiments. 
It is also possible to obtain direct measurements of flexibility by studying amide-protein exchange rates by NMR, 
but this does not provide detailed structural information on different structural states. 

Other experimental techniques can also provide information of flexibility. Small-angle X-ray scattering (SAXS) can determine a rough low-resolution shape of the molecules, but it is limited to a few hundred collected datasets. 
Different structures of (related) proteins solved by X-ray crystallography can shed light into the different conformational states of some proteins. 
However, only states that form stable crystal forms can be measured, limiting the types of flexibility that can be detected. Moreover, there is an imbalance in the PDB related to the abundance of pharmaceutically important proteins in complex with different ligands, or other factors. For example, the inactive state of kinases is largely underrepresented in the PDB compared to the active state \cite{schwarz2019modeling}. This may bias data-driven approaches while, in principle, without any extra-information about the context (post-translational modifications, bound ligand...etc), there is no reason why one state should be favoured over the other one. 
Cryo-EM can also provide information about multiple structural states as well as flexibility. Here, current methods are often limited to a fixed set of clearly distinguishable shapes/conformations present in the sample and selected during refinement.
Most of the flexibility information comes in the form of missing density, without any details about the flexible regions beyond the fact that they are flexible. However, we see the community moving toward the reconstruction of continuous structural heterogeneity, also using DL techniques \cite{zhong2021cryodrgn,punjani20213d,rosenbaum2021inferring}. Similar architectures, {\it i.e.} generative adversarial networks and variational auto-encoders, have also been used to generate protein backbones and produce smooth motions through linear interpolations in the latent space 
\cite{anand2019fully,eguchi2020ig}.

Finally, large collections of molecular dynamics (MD) trajectories \cite{meyer2010model,rodriguez2020gpcrmd} may be exploited toward protein flexibility learning.
However, today unbiased MD simulations are still too short (and likely too inaccurate) to sample large conformational changes. 
Therefore, learning from these simulations would be limited to small fluctuations around the starting structure. In fact, as we have seen in the recent CASP structure refinement studies, MD is only practical if additional restraints are applied to keep the structure near the initial conformation. Alternatively, instead of learning from MD trajectories, deep learning can be used to generate conformational ensembles obeying the Boltzmann distribution  \cite{noe2019boltzmann}. 

\subsection{Protein complexes and interactions}

Most proteins do not act alone. They function by interacting with other
proteins and molecules. 
Protein complexes come in different forms and shapes. A complex can consist of one or several types of molecules, contain anything from two to hundreds of different protein chains (as well as other macromolecules), and can have different degrees of symmetries. 
Experimentally,
the study of stable protein interactions can be carried out using various
techniques. While many of them only provide an estimate of the strength or probability of the interaction, structure determination methods, including crystallography and Cryo-EM electron microscopy, unveil the atomic-resolution details of the assemblies.

Co-evolutionary information can, in principle, also be used to extract
information about protein-protein interactions. Such strategy has been employed to predict bacterial complexes \cite{Schug20018738,Hopf25255213,Ovchinnikov24842992}, to single out pairs of interacting paralogs~\cite{Marmier31609984,Green33654096} and to gain insight into the interactomes of viruses, like HCV \cite{champeimont2016coevolution}, or bacteria, like {\it E. coli}~\cite{Cong31296772}. We do believe that this is the next area where end-to-end learning methods will make an impact. 
In contrast to the prediction of a single structure, one limitation here might be to detect a strong enough signal, since interactions across
protein (and domain) interfaces are less conserved than
intra-domain contacts \cite{Ghadie29112911}. 
Noticeably, the different types of assemblies have different specific properties and may require the development of different strategies. 

{\bf Homodimeric complexes} are special as the co-evolutionary signals
from a single protein describe both inter- and intra-protein
residue-residue interactions. Current methods
assume that only predicted interactions not satisfied within a single
protein (given some error margin) are potential inter-unit
connections although this is not always the
case~\cite{Quadir2020.11.09.373878}. 
It is also common that homomeric
protein complexes can adopt different quarternary forms, further
complicating the prediction, but we would expect that
extending AlphaFold2 to predict the structure of homodimers (and even
homo-multimers) should not be too difficult, at least as long as the multimeric formation is conserved within a family. What might prove to be more difficult is to identify the multimeric state of a protein without some type of experimental information. To the best of our knowledge, this problem is not yet studied.

{\bf Heterodimeric complexes} create a different challenge for
multiple sequence alignments. In short, here it is necessary to match
the exact pairs of interacting proteins from two lists of homologs. In rare cases, where there exists exactly one
homolog to each of the proteins in a genome, this is trivial. However,
many proteins have paralogs that might not all interact with each
other. Some paralogs might interact with the same protein and some
might not. One common approach is to identify the top hit in each
proteome --- but this is not always correct and it significantly reduces
the number of sequences in the MSA. 
In a small benchmark of 215 proteins \cite{Kundrotas28891124}, the structure for only a handful (5-10\%) of the complexes could be predicted correctly using a naive approach matching top hits from all genomes \cite{Lamb30796988}. 
Other methods trying to identify
the pairs might work better but are computationally
expensive \cite{Marmier31609984}. It is also possible that methods using unaligned sequences will provide a solution to this problem \cite{Weinstein2020.07.31.231381}.
Assuming that this problem can be solved, we do not see that there should exist any major obstacle to develop an end-to-end solution for the modeling of heterodimeric complexes.

{\bf Large molecular machines}, such as the ribosome, may represent the most challenging case. They typically perform very important functions in a cell. Their interaction networks may comprise very dense and stable subnetworks, and also parts where binary or ternary interactions are established at a given time-point. 
Recent Cryo-EM structures of large complexes have revealed that these machines often are dynamic with subunits coming
on and off. Clearly, we are still far away from being able to fully predict
their structure and dynamics. 

\added{In the few weeks since the release of the Alphafold2 source code, various groups have shown that the program in many cases can produce accurate assemblies, for most types of complexes. 
Different strategies seem to work for different types of complexes. 
First, many groups simply added a poly-G linker to join two (or more) chains and successfully docked both protein-protein and protein-peptide complexes. 
Later, it was discovered that the poly-G linker was not needed, as it was sufficient to just change the residue numbering. 
Interestingly, in some cases it seems not to be necessary to "merge" the alignments by matching orthologous pairs, instead, the alignments can just be added with gaps at the end - this strategy seems to be the best for homomeric complexes. 
Exact limits and optimal strategies for using Alphafold2 for docking will certainly be known in a short time - but we will have to wait for larger benchmarks and not just rely on anecdotal stories.
}

\subsection{Protein mutations and design}

Even one single-point mutation can have a dramatic effect on a protein's ability to fold and/or perform its function(s). In parallel to the evolution of CASP, the past few years have seen a significant improvement in the field of mutational outcome prediction. By leveraging the large amounts of available sequence data, several recent methods have achieved much higher accuracy than established popular approaches relying on a variety of sequence and structure-based features \cite{trinquier2021efficient,laine2019gemme,louie2018fitness,riesselman2018deep,flynn2017inference,hopf2017mutation,figliuzzi2016coevolutionary,mann2014fitness,ferguson2013translating}. 
These approaches make the estimation of the impact of every possible substitution at every position in a protein-coding genome computationally feasible \cite{frazer2020large}. They also hold great potential for guiding protein design and engineering \cite{russ2020evolution,biswas2021low}. The success of these methods lies in their ability to capture dependencies between protein residues either by explicitly estimating inter-residue (pairwise) couplings \cite{hopf2017mutation,figliuzzi2016coevolutionary} or by implicitly accounting for global sequence contexts \cite{laine2019gemme,riesselman2018deep}. In essence, the concepts at play are no different from those implemented for protein contact prediction, suggesting that mutational outcome prediction, protein structure prediction and protein design can be unified in a common theoretical framework extracting information from protein sequences \cite{trinquier2021efficient,Weinstein2020.07.31.231381,riesselman2018deep}.
Along this line, recent works have shown that NLP models pre-trained on millions of unlabelled protein sequences can be effectively fine-tuned with small amount of labelled data toward accurately predicting mutational effects as well as 3D contacts \cite{rives2021biological,biswas2021low,madani2020progen}.
Additionally, fully trained DNNs designed to predict inter-residue distances can be re-purposed to estimate the impact of mutations on the 3D structure toward guiding the generation of new sequences predicted to fold to new structures \cite{anishchenko2020novo}.

\section*{Acknowledgements}
SE was supported by a Stanford Bio-X Bowes Fellowship.
AE was funded by grants from the Swedish Natural Science Research
Council No. VR-NT 2016-03798, Swedish E-science Research Center and Swedish National Infrastructure for Computing. EL acknowledges the support of the French Agence Nationale de la Recherche (ANR) under reference ANR-17-CE12-0009. The founder had no role in study design,
data collection and analysis, decision to publish, or preparation of
the manuscript. 
The authors thank Kliment Olechnovič from Vilnius University for his help with illustrating Voronoi cells and proof-reading the manuscript,
and Bowen Jing for his feedback on the manuscript. 

\section*{Conflict of interest}
The authors declare no conflict of interest. 
%

\printendnotes




\end{document}